\documentclass[aps,twocolumn,dvipdfm]{revtex4}

\usepackage[latin1]{inputenc}
\usepackage[dvipdfm]{graphicx}
\usepackage{amsmath}
\usepackage{color,textcomp}

\newcommand{\ket}[1]{\left| #1 \right>}
\newcommand{\Dge}{\Delta_{ge}}

\newcommand{\avg}[1]{\bar #1}
\newcommand{\EE}{\mathcal{E}}
\newcommand{\acro}{ISG}

\begin{document}
\title{Interlaced spin grating for optical wave filtering}
\author{H. Linget$^1$, T. Chaneli\`ere$^1$, J.-L. Le Gou\"et$^1$, P. Berger$^2$, L. Morvan$^2$, and A. Louchet-Chauvet$^1$}
\affiliation{$^1$Laboratoire Aim\'{e} Cotton, CNRS UPR3321, Univ. Paris Sud, B\^at. 505, campus universitaire, 91405 Orsay, France\\
$^2$Thales Research and Technology, 1 Avenue Augustin Fresnel, 91767 Palaiseau, France}
\email{anne.chauvet@u-psud.fr}

\begin{abstract}
Interlaced Spin Grating is a scheme for the preparation of spectro-spatial periodic absorption gratings in a inhomogeneously broadened absorption profile. It relies on the optical pumping of atoms in a nearby long-lived ground state sublevel. The scheme takes advantage of the sublevel proximity to build large contrast gratings with unlimited bandwidth and preserved average optical depth. It is particularly suited to Tm-doped crystals in the context of classical and quantum signal processing. In this paper, we study the optical pumping dynamics at play in an Interlaced Spin Grating and describe the corresponding absorption profile shape in an optically thick atomic ensemble. We show that, in Tm:YAG, the diffraction efficiency of such a grating can reach $18.3\%$ in the small angle, and $11.6\%$ in the large angle configuration when the excitation is made of simple pulse pairs, considerably outperforming conventional gratings.

\end{abstract}
\date{\today}
%\ocis{(000.0000) General. *********}
%%%%%%%%%%%%%%%%%%%%%%%%%%  body  %%%%%%%%%%%%
\maketitle

\section{Introduction}
Thanks to their uncommon spectroscopic properties, rare-earth ion-doped crystals (REIC) have been proposed for a large variety of signal processing applications, stretching from classical signal processing to quantum memories.

Most classical signal processing architectures using REIC are based on the creation of a grating imprinted in their absorption profile~\cite{babbitt1998review}. This grating can be either spectral or spectro-spatial. Angle of arrival estimation~\cite{barber2010angle} and time reversal~\cite{linget2013} are based on spectral gratings. Analog-to-digital conversion~\cite{babbitt2007adc}, detection of correlations~\cite{merkel2000,harris2000,schlottau2004}, lidar range Doppler imaging~\cite{li2006lidar}, true-time delay~\cite{tian2001,reibel2004ttd} or spectral analysis~\cite{lavielle2003} use spectro-spatial gratings.
Many of these architectures have been demonstrated in Tm:YAG, a REIC that provides multi-GHz bandwidth, sub-kHz resolution limit, and accessibility to diode lasers altogether. However, the diffraction efficiencies did not exceed $1$\% in the experimental demonstrations.
%because only a low-contrast grating can exist over a large bandwidth (while avoiding massive excitation of the atoms????????).

The atomic frequency comb (AFC)~\cite{afzelius2009}, one of the many quantum memory protocols developed specifically for REIC, is also based on a spectral grating (or spectro-spatial grating in the case of the spectro-spatial atomic comb (S2AC)~\cite{tian2013}). Surprisingly enough, diffraction efficiencies up to $17$\%~\cite{bonarota2010efficiency} in Tm:YAG and $35$\% in Pr:YSO~\cite{amari2010} have been observed with AFC. Reaching such efficiencies requires the use of long-lived ground state sublevels for storage instead of the metastable level used in the classical processing experiments.
Yet the presence of closeby sublevels within the inhomogeneous line leads to grating replicas at other frequencies, limiting the grating bandwidth to the smallest ground or excited state splitting (typically a few MHz in Tm:YAG or Pr:YSO).
%Only Kramers ions such as Nd allow for a $100$~MHz-wide grating due to their GHz ground state splitting~\cite{usmani2010mapping}.

Nevertheless, a spectral or spectro-spatial grating in a REIC with closeby ground state sublevels can exist over a broad bandwidth, by transferring atoms between the ground state sublevels from bright fringes to dark fringes. The grating period must be adjusted so that the grating replicas add constructively with the original one. This way, the total grating bandwidth can considerably exceed the ground state splitting. %Because the atoms are not removed but rearranged within the grating bandwidth, the average optical depth is preserved.
The absorbing and transparent fringes are intrinsically and unavoidably linked together since they mutually feed each other. In order to emphasize this inseparability, we propose to name the scheme "Interlaced Spin Grating" (\acro).
It has been used for example by Saglamyurek \emph{et al.}~\cite{saglamyurek2011} in a Ti:Tm:LiNbO$_3$ optical waveguide, where an AFC is created over a $5$~GHz bandwidth, exceeding the ground state splitting by a factor of 70, or by Bonarota \emph{et al.}~\cite{bonarota2011highly} in Tm:YAG, where this factor reaches $370$. In these two demonstrations however, the diffraction efficiency did not exceed $2$\%.

In the applications we have in mind, the active material behaves as a programmed filter. This filter has to process weak signals and to offer a large dynamic range. In addition the signals may reach the filter at random times and we aim at $100\%$ interception efficiency, which requires continuous operation capability. Only a passive filter may satisfy those conditions. Filtering through an inverted medium~\cite{szabo1984,dasilva1993,crozatier2005} may offer large diffraction efficiency but is not consistent with the above requirements. Medium inversion entails spontaneous emission and even amplified spontaneous emission. Besides, continuously refreshed inversion would require high intensity continuous illumination, which is source of heating and is not compatible with operation at cryogenic temperature. Finally, massive excitation of the active medium gives rise to complex relaxation processes, such as energy migration, that deeply alter the programmed function~\cite{french1992energy}.

In this paper, we explore the mechanisms at play in an ISG and identify the fundamental limits to its efficiency. Sec.~\ref{sec:OP} is devoted to the comparison of different optical pumping mechanisms used to create absorption gratings. The \acro\ mechanism is explicited. The limits to the engraving power are highlighted in the different optical pumping schemes. In Sec.~\ref{sec:deep} we describe the effect of propagation on the engraving beams, and its consequences on the engraved grating shape and contrast. Numerical simulations are presented. We derive the corresponding diffraction efficiencies in various configurations, from the pure spectral grating with co-propagating engraving beams, to the spectro-spatial grating. We conclude this paper by presenting an experimental verification of these results in Sec.~\ref{sec:exp}.

%show that the diffraction efficiencies in Tm are below $2$~\% when using the metastable state as storage level. On the contrary, the \acro\ scheme allows for an enhancement of the grating contrast, together with quasi-optimal grating shape, resulting in a much larger diffraction efficiency.

\section{Optical pumping}
\label{sec:OP}
In this section we compare the optical pumping mechanisms in the conventional scheme where the atoms are stored in a level outside the operation bandwidth, and in the \acro\ scheme where the atoms are stored in a long-lived sublevel within the operation bandwidth. Finally, we discuss the use of \acro\ in Tm-doped crystals.

\subsection{Standard optical pumping}
\subsubsection{Optical pumping mechanism}
\label{sec:OP3level}

\begin{figure}[t]
\centering
\includegraphics[width=5.5cm]{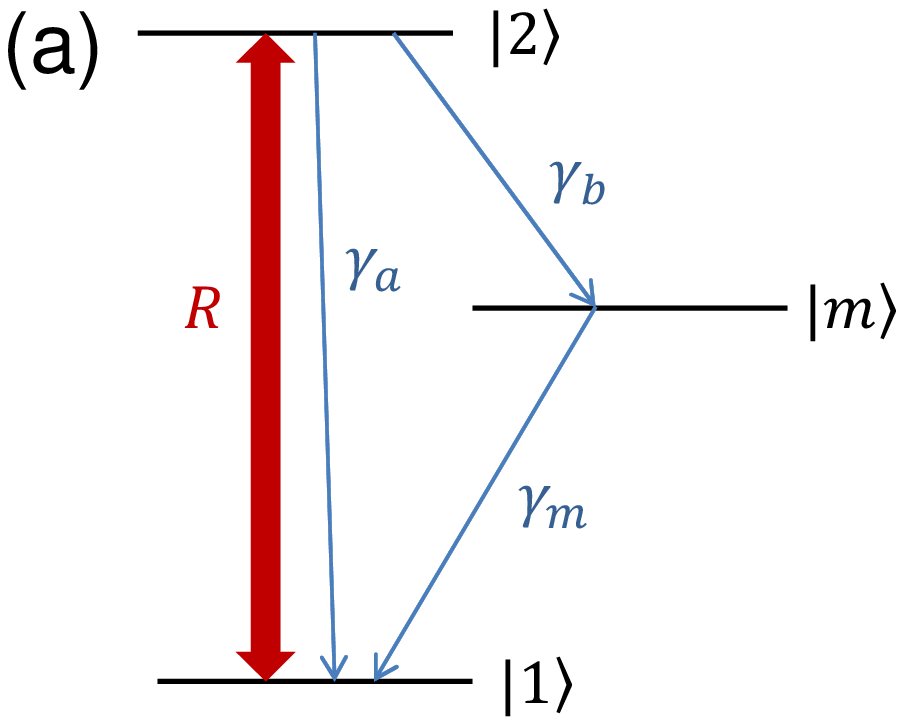}
\includegraphics[width=2.8cm]{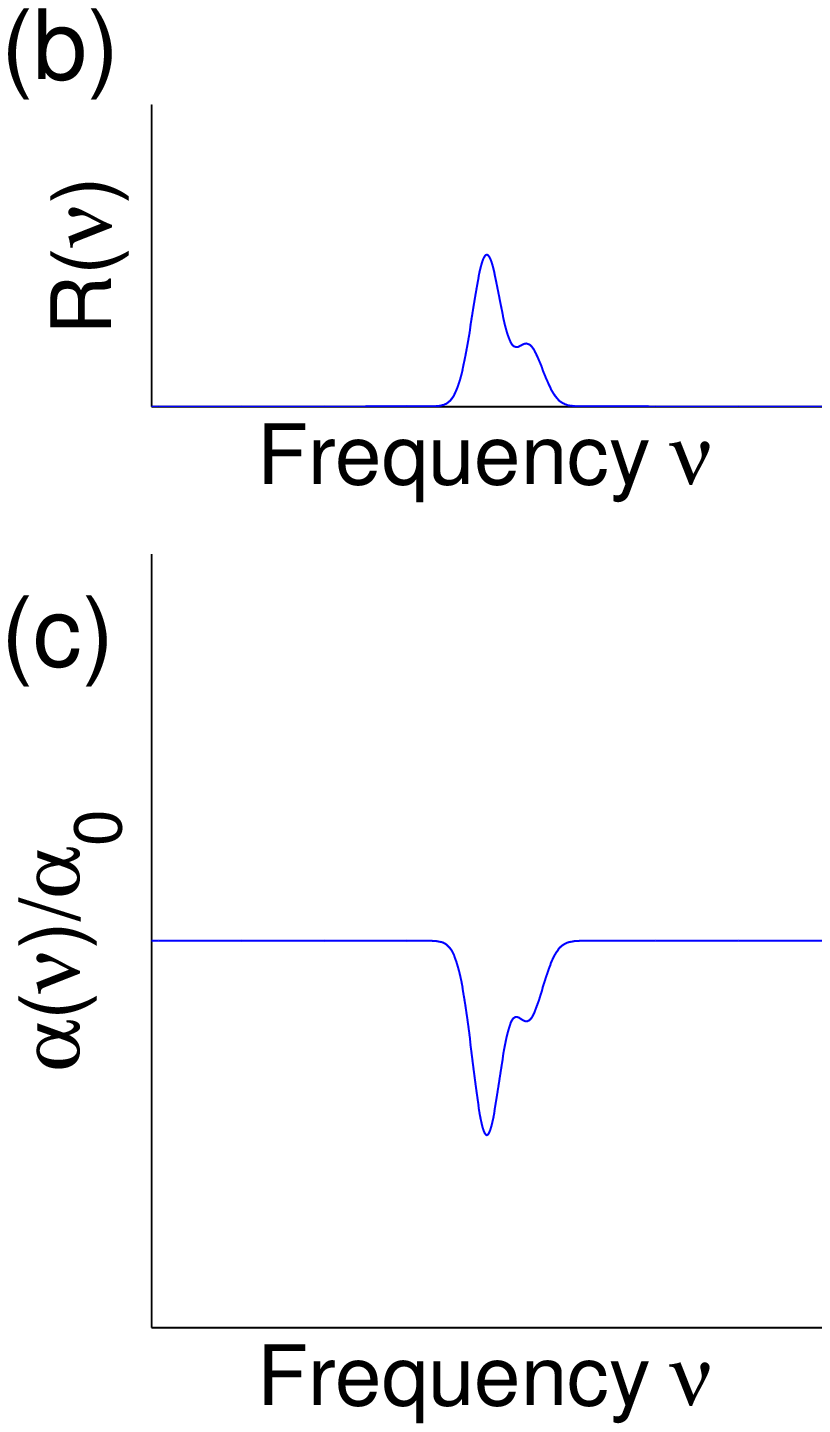}
\caption{(Color online) (a): 3-level system for standard optical pumping. (b) steady-state excitation $R(\nu)$. (c) Shape of the resulting absorption profile.}
\label{fig:3level}
\end{figure}

The simplest level scheme where optical pumping can be observed is given in Fig.~\ref{fig:3level}(a). It consists in three levels $\ket{1}$, $\ket{2}$ and $\ket{m}$, such that the excitation pumps atoms from $\ket{1}$ to $\ket{2}$, and the intermediate, metastable level $\ket{m}$, lies outside the pumping spectrum.
The atoms in the upper state $\ket{2}$ can decay back to the ground state $\ket{1}$ with rate $\gamma_a$ or to state $\ket{m}$ with rate $\gamma_b$. $\ket{m}$ is considered as metastable because its decay rate $\gamma_m$ to the ground state is much smaller than the total decay rate from the upper state $\gamma_e=\gamma_a+\gamma_b$.

In the case where the excitation is weak, $\emph{ie}$ when the Fourier transform $\tilde{\Omega}(\nu)$ of the Rabi frequency $\Omega(t)$ obeys
\begin{equation}
| \tilde\Omega(\nu) |^2 \ll 1,
\label{eq:weak}
\end{equation}
the transition is not saturated. The excitation along $\ket{1} \rightarrow \ket{2}$ can be described through the average pumping rate $R(\nu)$ and the optical Bloch equations reduce to rate equations. The stationary population difference along this transition is given by
\begin{equation}
\Delta n_{12}=n_1-n_2=\frac{1}{1+\zeta r}
\label{eq:n3}
\end{equation}
where $n_1$ (resp., $n_2$) is the proportion of atoms in  state $\ket{1}$ (resp. $\ket{2}$) and $\zeta=(\gamma_b+2\gamma_m)/\gamma_e$. The reduced pumping rate is defined as $r=R/\gamma_{m}$. The absorption coefficient is given by
\begin{equation}
\alpha=\alpha_0 \Delta n_{12}
\label{eq:alpha3}
\end{equation}
where $\alpha_0$ is the initial absorption.

We assume the homogeneous linewidth to be much smaller than the resolution of $\tilde\Omega(\nu)$. Its convolution effect is then neglected. If the transition is not saturated, then the average pumping rate over duration $T$ reads as $R(\nu)=|\tilde\Omega(\nu)|^2/(4T)$. Because the atomic absorption lines are spread over the inhomogeneous absorption profile, the population difference $\Delta n_{12}(\nu)$ is modified according to Eq.~\ref{eq:n3}, where $R$ is replaced with $R(\nu)$.

While $\zeta r \ll 1$, the modification of the absorption profile imitates the pumping spectrum: $\alpha(\nu)\simeq\alpha_0\left[1-\zeta r(\nu)\right]$, as illustrated in Fig.~\ref{fig:metastable_profil_entree}(b). As the pumping rate increases, the engraved structure depth grows, until saturation broadens the absorption structures. The average absorption decreases to zero as the pumping rate increases, since atoms accumulate in state $\ket{m}$, outside the grating bandwidth.

\subsubsection{Construction of a spectro-spatial grating}
\begin{figure}[t]
\includegraphics[width=7.5cm]{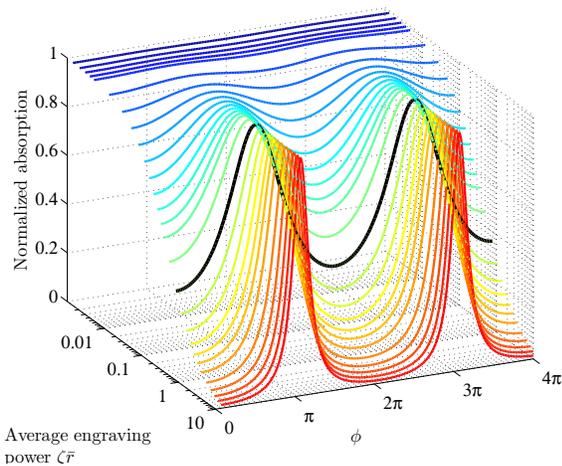}
\caption{(Color online) Absorption grating in the standard optical pumping scheme, given a sinusoidal spectro-spatial pumping rate, assuming a weak excitation as specified in Eq.~\ref{eq:weakR}. Black line: spectral grating obtained with the strongest average pumping rate as allowed by Eq.~\ref{eq:weakR} in the specific case of Tm:YAG.}
\label{fig:metastable_profil_entree}
\end{figure}

In the discussion above we have implicitly considered the excitation to be continuous, but a spectro-spatial grating is generally created with pulse pairs. Let us consider two short rectangular pulses with equal power and duration, separated by a delay $\tau$ and with respective wavevectors $\textbf{k}_0$ and $\textbf{k}_1$. The $z$ axis is defined parallel to $\textbf{k}_0+\textbf{k}_1$. The reduced pumping rate then depends on the frequency $\nu$ and transverse position $\textbf{x}$ in the crystal through the spectro-spatial phase $\phi$:
\begin{equation}
r(\phi)=\avg{r}\left[1+\cos(\phi)\right] \textrm{with } \phi=2\pi\nu \tau+\textbf{K}\cdot\textbf{x}
\end{equation}
where $\textbf{K}=\textbf{k}_1-\textbf{k}_0$ and $\avg{r}$ is the average value of the reduced pumping rate $r$ over the excitation spectrum.  Fig.~\ref{fig:metastable_profil_entree} shows the shape of the absorption coefficient $\alpha(\phi)$ for different average pumping powers. It appears that the absorption profile can be sinusoidal only for low engraving power. A larger pumping rate makes the contrast larger but the profile departs from the sinusoidal shape, getting closer to a comb-like shape. We define a specific contrast parameter as: \begin{equation}
c=\frac{\alpha_{max}-\alpha_{min}}{\alpha_0}
\label{eq:contrast}
\end{equation}
where $\alpha_0$ is the initial absorption coefficient.
We observe that $c$ starts from $0$ and tends to $1$ as the engraving power is increased.

Considering that during a time interval $T$ we send one pulse pair, the average pumping rate over $T$ reads $R=A^2(1+\cos\phi)/(2T)$ where $A$ is the area of an individual pulse. The time interval $T$ should be  long enough to let the atoms relax to lower states ($T\geq \frac{2}{\gamma_e}$). Therefore the weak field condition given in Eq.~\ref{eq:weak} becomes:
\begin{equation}
R \ll \frac{\gamma_e}{2}
\label{eq:weakR}
\end{equation}

In Tm-doped YAG, the decay rates are $\gamma_a=\frac14 \gamma_e$, $\gamma_b=\frac34\gamma_e$, with $\gamma_e=1/800\mu$s and $\gamma_{m}=1/10$ms~\cite{deseze2005}, hence $\zeta=0.91$. With these parameters, the weak field condition translates as $\zeta \avg{r}\ll 6$.
The deepest grating allowed by this condition with the Tm:YAG parameters is plotted as a black line in Fig.~\ref{fig:metastable_profil_entree}. Its contrast (defined in Eq.~\ref{eq:contrast}) is equal to $0.63$.

\subsection{Optical pumping with a sublevel structure}

\subsubsection{Optical pumping mechanism}
\begin{figure}[t]
\includegraphics[width=5cm]{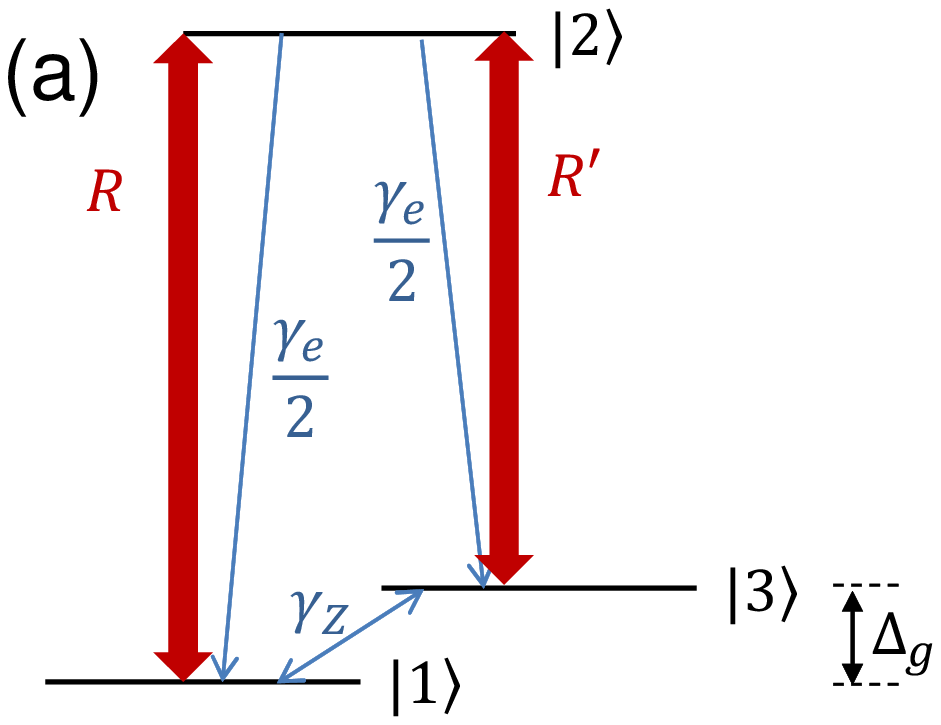}
\includegraphics[width=2.8cm]{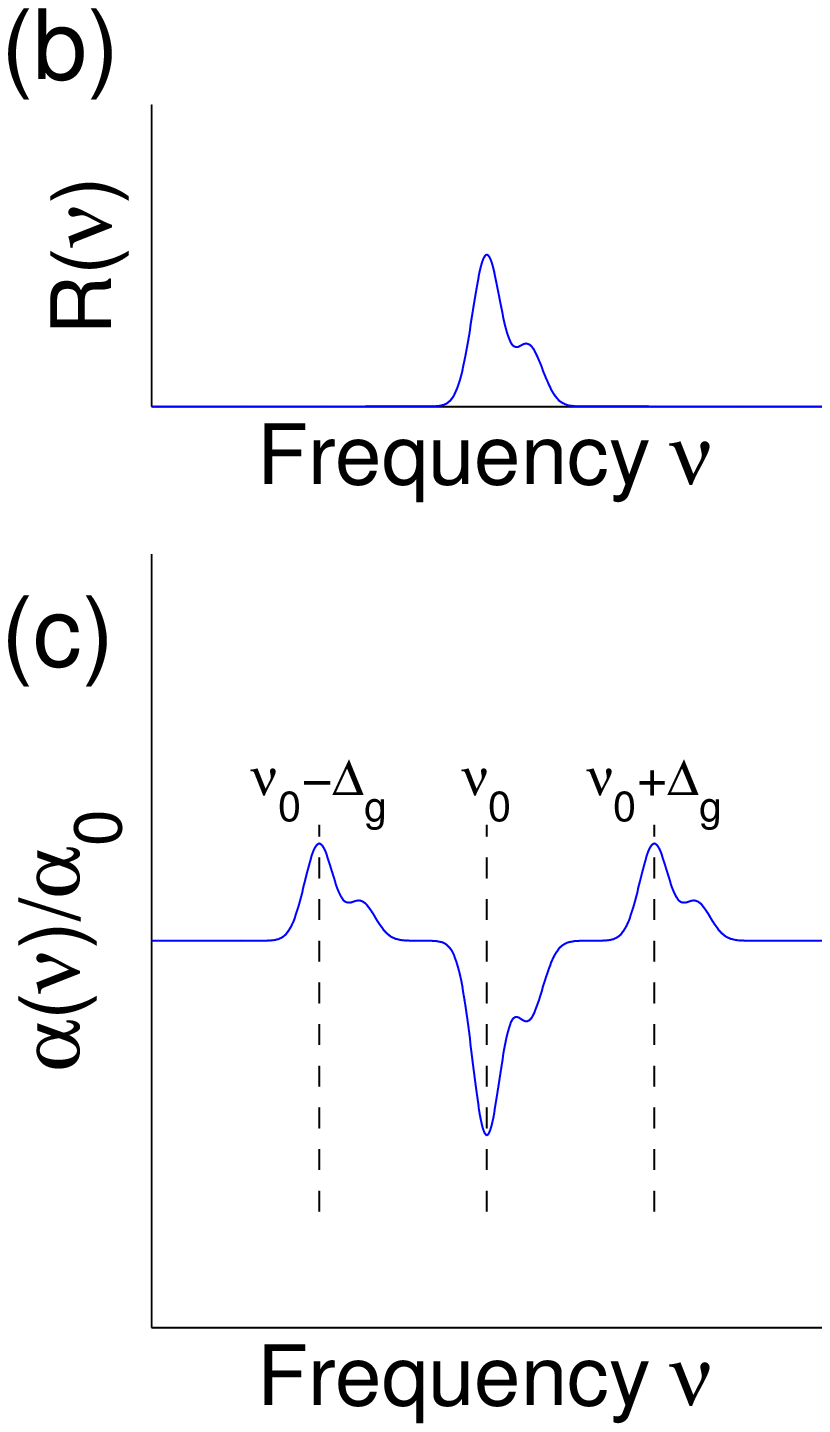}
\caption{(Color online) (a) 3-level system with a sublevel structure in the ground state. (b) steady-state excitation $R(\nu)$. (c) Shape of the resulting absorption profile, with side structures due to the closeby ground state sublevels.}
\label{fig:3levelZ}
\end{figure}

Let us now consider a three-level $\Lambda$-type system where two spin sublevels $\ket{1}$ and $\ket{3}$ of a ground state are connected via optical transitions to one upper level $\ket{2}$ (see Fig.~\ref{fig:3levelZ}). Now both ground state sublevels are long-lived and can be used as storage levels. They are connected via an optical transition to the upper state. Again, we consider the weak field situation (Eq.~\ref{eq:weakR}), such that system evolution can be described by rate equations. The frequency difference $\Delta_g$ between the two lower states is assumed to be smaller than the inhomogeneous linewidth of the optical transition. This way, the atoms are simultaneously excited along their two optical transitions with pumping rates $R$ and $R'$. We assume that the transition probabilities are equal, so that $R'(\nu)=R(\nu-\Delta_g)$. The total decay rate from the upper state $\gamma_e$ is assumed to be much larger than the ground state sublevel relaxation rate $\gamma_Z$. The steady-state solution to the rate equations yields the following population differences:
\begin{eqnarray}
\Delta n_{12}=\frac{\frac 12+\xi r'}{1+\xi(r+r')}
\label{eq:n3Za}\\
\Delta n_{32}=\frac{\frac 12+\xi r}{1+\xi(r+r')}
\label{eq:n3Zb}
\end{eqnarray}
where the reduced pumping rates are defined as $r=R/\gamma_e$ and $r'=R'/\gamma_e$, and $\xi=\gamma_e/(2\gamma_Z)$.
The resulting absorption is obtained by combining the two quantities above:
\begin{equation}
\alpha(\nu)=\alpha_0 [\Delta n_{12}(\nu) + \Delta n_{32}(\nu+\Delta_g)]
\label{eq:alpha3Z}
\end{equation}
where $\alpha_0$ is the initial absorption coefficient.
This gives rise to a depletion of the absorption profile according to the pumping rate $R(\nu)$, but also to inverted replicas of the engraved structure at $\pm \Delta_g$ where the absorption is increased, as shown in Fig.~\ref{fig:3levelZ}(b).

\subsubsection{Construction of a spectro-spatial grating }
\begin{figure}[t]
\includegraphics[width=8cm]{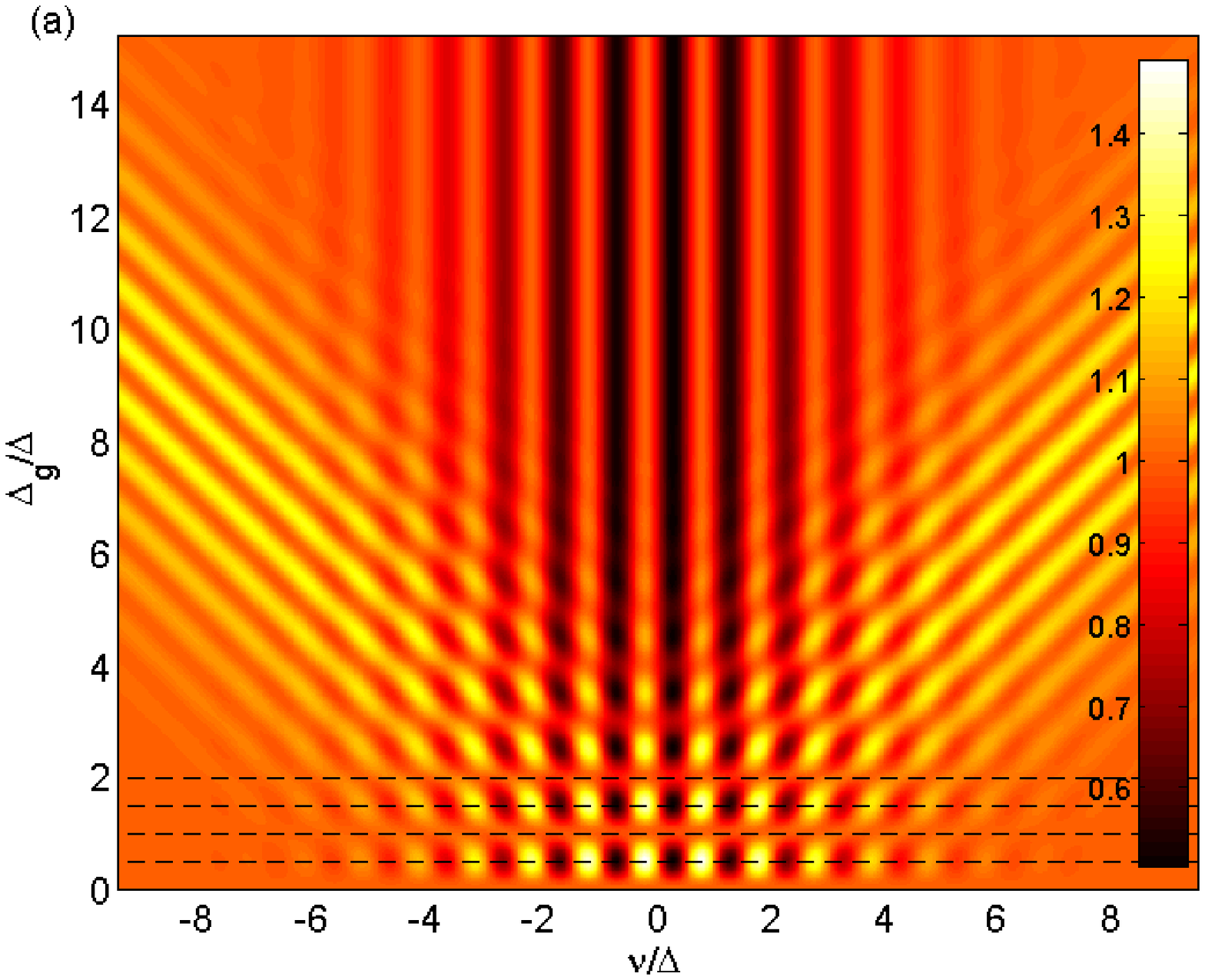}
\includegraphics[width=7cm]{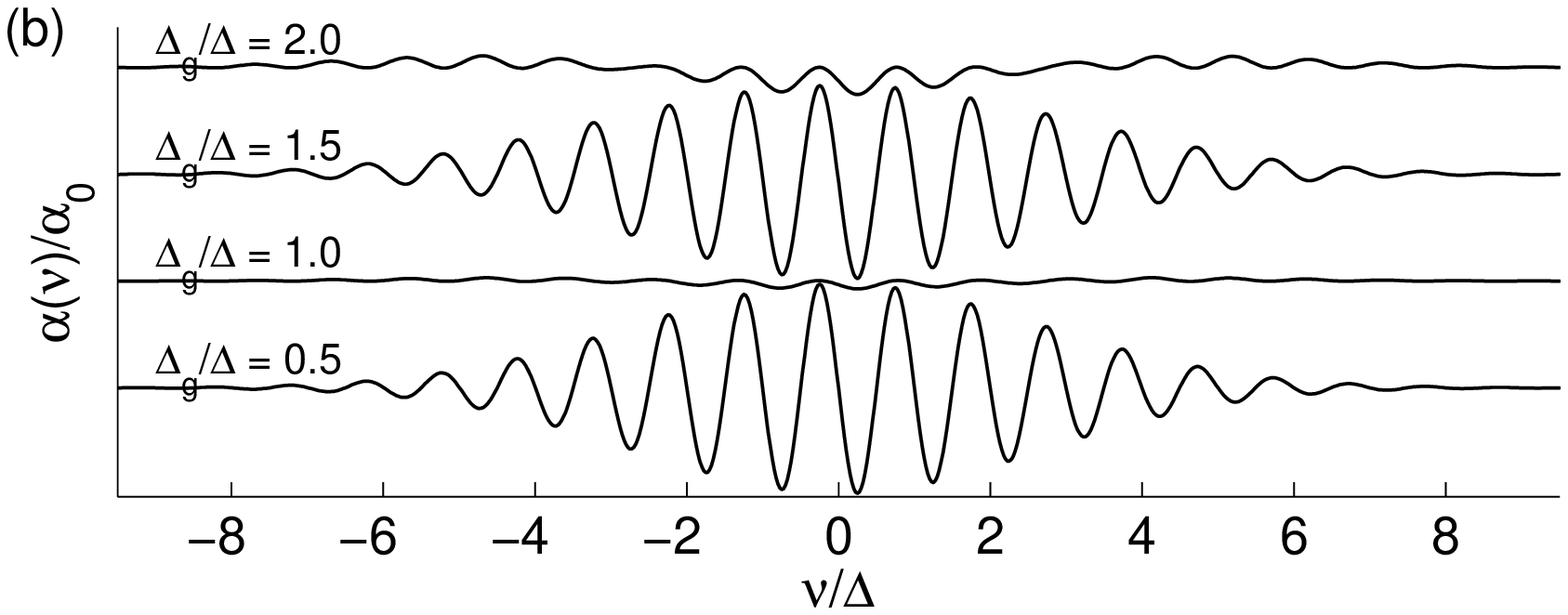}\rule{.5cm}{0cm}
\caption{(Color online) (a) Normalized absorption profile $\alpha(\nu)/\alpha_0$ under steady-state excitation $R(\nu)$, for different ground state splittings $\Delta_g$. The spectrally-dependent pumping rate $R(\nu)$ spectrum corresponds to excitation by two consecutive gaussian pulses. (b) Absorption profiles when $\Delta_g$ is smaller than the excitation spectral width, for 4 values of ratio $\Delta_g/\Delta$, where $\Delta$ is the spectral period of pumping rate $R(\nu)$. The curves are vertically offset for clarity.}
\label{fig:ISG}
\end{figure}

%\label{sec:ISG}
Consider the situation where the excitation spectrum has a spectrally periodic structure around central frequency $\nu_0$. While the splitting $\Delta_g$ is larger than the excitation spectral width, the three components at $\nu_0$, $\nu_0+\Delta_g$ and $\nu_0-\Delta_g$ are well separated in the absorption spectrum [see Fig.~\ref{fig:ISG}]. Alternatively, when the frequency difference $\Delta_g$ is smaller than the excitation spectral width, the replicas overlap with the central structure, Xand the relative values of $\Delta_g$ and the grating period $\Delta=1/\tau$ become crucial. For example, when $\Delta_g$ is an integer multiple of $\Delta$, the different contributions almost cancel out and the grating contrast is weak. But when $\Delta_g$ is a half-integer multiple of $\Delta$, the three contributions add constructively, giving rise to a grating with an enhanced contrast. The resulting absorption grating is made from the superposition of three overlapping gratings due to periodic storage in both ground state spin sublevels. Because these three gratings are inseparable and overlapping, and because the atoms are kept in the two ground state spin sublevels, we name this the "Interlaced Spin Grating" (\acro).
It is related to the technique used by G. Pichler~\emph{et al.}~\cite{aumiler2005,ban2006} to modulate the absorption profile of a rubidium vapor, with the help of a femtosecond pulse-train excitation with a spectral width ($10$~nm) several orders of magnitude larger than the Doppler broadening (about $500$~MHz).

\begin{figure}[t]
\includegraphics[width=7.5cm]{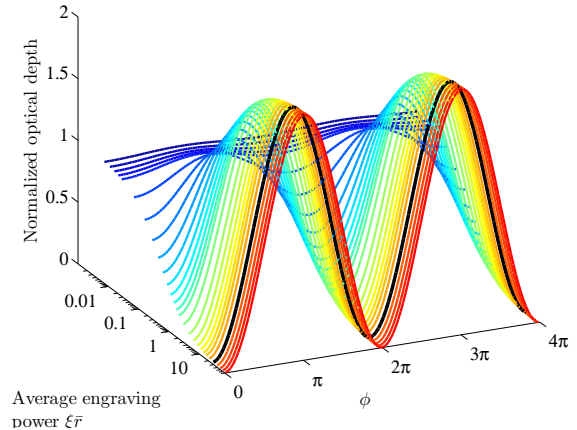}
\caption{(Color online) Absorption grating in the \acro\ scheme, given a sinusoidal spectro-spatial pumping rate, assuming a weak excitation as specified in Eq.~\ref{eq:weakR}. Black line: \acro\ grating obtained with the strongest average pumping rate as allowed by Eqs.~\ref{eq:weakR} and~\ref{eq:weak3Z} in the specific case of Tm:YAG.}
\label{fig:zeeman3_profil_entree}
\end{figure}

Let us consider an infinitely broad, spectro-spatially periodic pumping rate $r(\phi)=\avg{r}\left[1+\cos(\phi)\right]$, where $\phi= 2\pi\nu \tau + \textbf{K} \cdot \textbf{x}$ is the spectro-spatial phase.
We adjust $\Delta_g \tau=1/2$ to make \acro\ enhancement possible. Indeed, $r'(\phi)=\avg{r}\left[1-\cos(\phi)\right]$, in perfect antiphase with $r(\phi)$. The absorption profile $\alpha(\phi)$ reads as:
\begin{equation}
\alpha(\phi)=\alpha_0\left( 1-\frac{2\xi\avg{r}}{1+2\xi\avg{r}}\cos\phi\right)
\label{eq:alpha3cos}
\end{equation}
The absorption profile is therefore sinusoidal, with its average value being kept constant: the absorption is reduced in the bright excitation regions, and enhanced in the dark regions, approaching \emph{twice} the initial absorption. As the engraving power is increased, the grating contrast (as defined in Eq.~\ref{eq:contrast}) grows asymptotically from $0$ to $2$, but no distortion occurs (see Fig.~\ref{fig:zeeman3_profil_entree}). Therefore, with \acro\ and a sinusoidal excitation spectrum with an arbitrarily broad envelope, one can create a sinusoidal absorption profile with a contrast close to $2$, which was impossible in the standard optical pumping scheme. The bandwidth of such a grating is no longer limited by the atomic level splitting.

\acro\ is not limited to level systems strictly corresponding to Fig.~\ref{fig:3levelZ}. Its implementation is rather complex though in systems where the ground state contains three levels or more, because there are more than two storage states. This is the case for Eu$^{3+}$- or Pr$^{3+}$-doped crystals.
On the other hand, the \acro\ scheme can be applied in a straightforward way to REIC with only two ground states. This is true for instance for Kramers ions such as Er$^{3+}$ or Nd$^{3+}$ under magnetic field, where the Kramers doublets are split into two sublevels, or for Tm$^{3+}$ under magnetic field, where the nuclear Zeeman effect splits the electronic levels. Nevertheless, the Tm level scheme is slightly more complex, as we will see in the following.

\subsection{Interlaced Spin Grating in Tm-doped crystals}
\label{sec:ISGTm}
\begin{figure}[t]
\includegraphics[width=4.5cm]{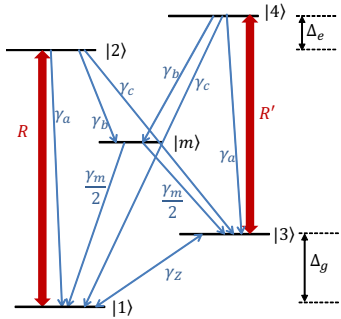}
\caption{(Color online) Thulium 5-level system and notations. $\Dge=\Delta_g-\Delta_e$ is defined as the frequency offset between the two optical transitions.}
\label{fig:5level}
\end{figure}

The thulium level system consists of 5 levels: 2 sublevels of the ground and excited states split by the nuclear Zeeman effect, and a metastable state (see Fig.~\ref{fig:5level}).
Optical transitions occur along $\ket{1} \rightarrow \ket{2}$ and $\ket{3} \rightarrow \ket{4}$ with equal oscillator strength. Their frequency offset $\Dge=\Delta_g-\Delta_e$ is orders of magnitude smaller than the inhomogeneous broadening of the optical transition. The other transitions ($\ket{1} \rightarrow \ket{4}$ and $\ket{3} \rightarrow \ket{2}$) can be weakly allowed in Tm:YAG, but only for very specific magnetic field orientations~\cite{deseze2006}. In the following we will consider these weak transitions as forbidden.

The decay mechanism from the excited state occurs both as a direct radiative decay to the ground state, and via the metastable state. The direct relaxation can be considered as completely spin-preserving (\emph{ie} $\gamma_c=0$). The indirect relaxation via the metastable state probably involves some spin mixing. For the sake of simplicity, we suppose that the atoms in the metastable state decay equally to both ground states, irrespective of their nuclear spin.

%
%occurs for the largest part as a non-radiative process, and is therefore spin-insensitive. This is why we consider a single metastable state $\ket{m}$, overlooking its nuclear Zeeman structure, and assume that the atoms relax from this metastable state equally to the two ground state sublevels.
%
This system can be seen as two independent two-level systems, with a shared decay channel via the intermediate metastable state $\ket{m}$. Excessive engraving power may lead to a significant fraction of atoms accumulating in $\ket{m}$, resulting in a lower average absorption, and in a non-stationary state when the engraving beams are stopped, due to the population decay from $\ket{m}$. We derive the condition on $R$ ensuring that all atoms are in the two ground states:
\begin{equation}
R\ll \gamma_m \gamma_e/\gamma_b
\label{eq:weak3Z}
\end{equation}
This condition must be satisfied together with Eq.~\ref{eq:weakR} ensuring the non-saturation of the optical transition.

Considering that the total excited state population decay rate $\gamma_e=\gamma_a+\gamma_b+\gamma_c$ is much larger than the ground state relaxation rate $\gamma_Z$, the steady-state solution to the rate equations is given by:
\begin{eqnarray}
\Delta n_{12}=\frac{\frac 12+\xi r'}{1+\xi(r+r')}
\label{eq:n5Za}\\
\Delta n_{34}=\frac{\frac 12+\xi r}{1+\xi(r+r')}
\label{eq:n5Zb}
\end{eqnarray}
with $r=\frac{R}{\gamma_e}$, $r'=\frac{R'}{\gamma_e}$ where $R'(\nu)=R(\nu-\Delta_{ge})$, and $\xi = \frac{\gamma_b/2+\gamma_c} {2\gamma_Z}$.
The resulting absorption is obtained by combining the two quantities above:
\begin{equation}
\alpha(\nu)=\alpha_0 [\Delta n_{12}(\nu) + \Delta n_{34}(\nu+\Delta_{ge})]
\label{eq:alpha5<Z}
\end{equation}
where $\alpha_0$ is the initial absorption coefficient.
In the case where the splitting $\Dge$ matches a half integer multiple of the grating's spectral period, we can derive Eq.~\ref{eq:alpha3cos}, just like in the simple 3-level system.

In Tm:YAG, the decay rates $\gamma_a$, $\gamma_b$, $\gamma_e$ and $\gamma_{m}$ have been given in Sec.~\ref{sec:OP3level} and we take $\gamma_c=0$ and $\gamma_Z=1/5$s.
If the pumping rate obeys Eq.~\ref{eq:weak3Z} ensuring that no atoms remain in the metastable state, then the weak field condition (Eq.~\ref{eq:weakR}) is immediately satisfied as well, since the ratio $\gamma_m /\gamma_b$ is much smaller than $1$.
More specifically, these two conditions are satisfied when $\xi\avg{r}\leq30$. The grating with a maximum contrast allowed by these two conditions ($\xi\avg{r}=30$) is plotted as a black line in Fig.~\ref{fig:zeeman3_profil_entree}. Its contrast (as defined in Eq.~\ref{eq:contrast}) is equal to $c=1.97$.
%$\gamma_a=\frac14\gamma_e$, $\gamma_b=\frac34\gamma_e$, $\gamma_a=\frac14\gamma_e$,

Therefore a broadband, perfectly sinusoidal grating can be created by accumulating pulse pairs in Tm-doped crystals.

\section{Engraving a grating in an optically thick medium}
\label{sec:deep}
In the previous section, we have overlooked the propagation of the engraving beams  in the atomic medium. In fact, the absorption profile presented above is only valid at the front of the medium.
Since we consider a steady-state accumulation regime, the engraving pulses propagate through the absorption grating they are creating, and undergo diffraction: higher order fields are emitted. The pumping rate evolves as the engraving fields penetrate in the optically thick medium and as a consequence, so does the absorption.

In a spectro-spatial grating, diffraction occurs both in the time domain as a delay, and in the space domain as an angular deviation.
We define the $n$-th order of diffraction as the pulse emitted with a delay $\tau_n=n \tau$ along wavevector $\textbf{k}_n=\textbf{k}_0+n\textbf{K}$. The $n$-th order of diffraction can be radiated only if the dipoles are phase-matched with the radiated field over the whole medium depth $L$. The corresponding phase matching condition reads as $| |\textbf{k}_n|-|\textbf{k}_0| | L \ll \pi$, which translates as:
\begin{equation}
n(n-1)\frac{|\textbf{K}|^2}{|\textbf{k}_0|}L\ll \pi
\label{eq:PMC}
\end{equation}

In the following, we focus on the two extreme cases: the configuration where the angle $\theta$ between $\textbf{k}_0$ and $\textbf{k}_1$ is so small ($\theta\ll \sqrt{\frac{\lambda}{2L}}$) that the phase-matching condition is satisfied for all significant diffraction orders, and the configuration where the angle $\theta$ is so large ($\theta\gg\sqrt{\frac{\lambda}{2L}}$) that only $n=0$ and $n=1$ can exist. The small angle situation includes the collinear situation where the grating is purely spectral ($\textbf{K}=0$).
A detailed description of the propagation of the engraving beams is given in the appendix.

\begin{figure}[t]
\includegraphics[width=7.5cm]{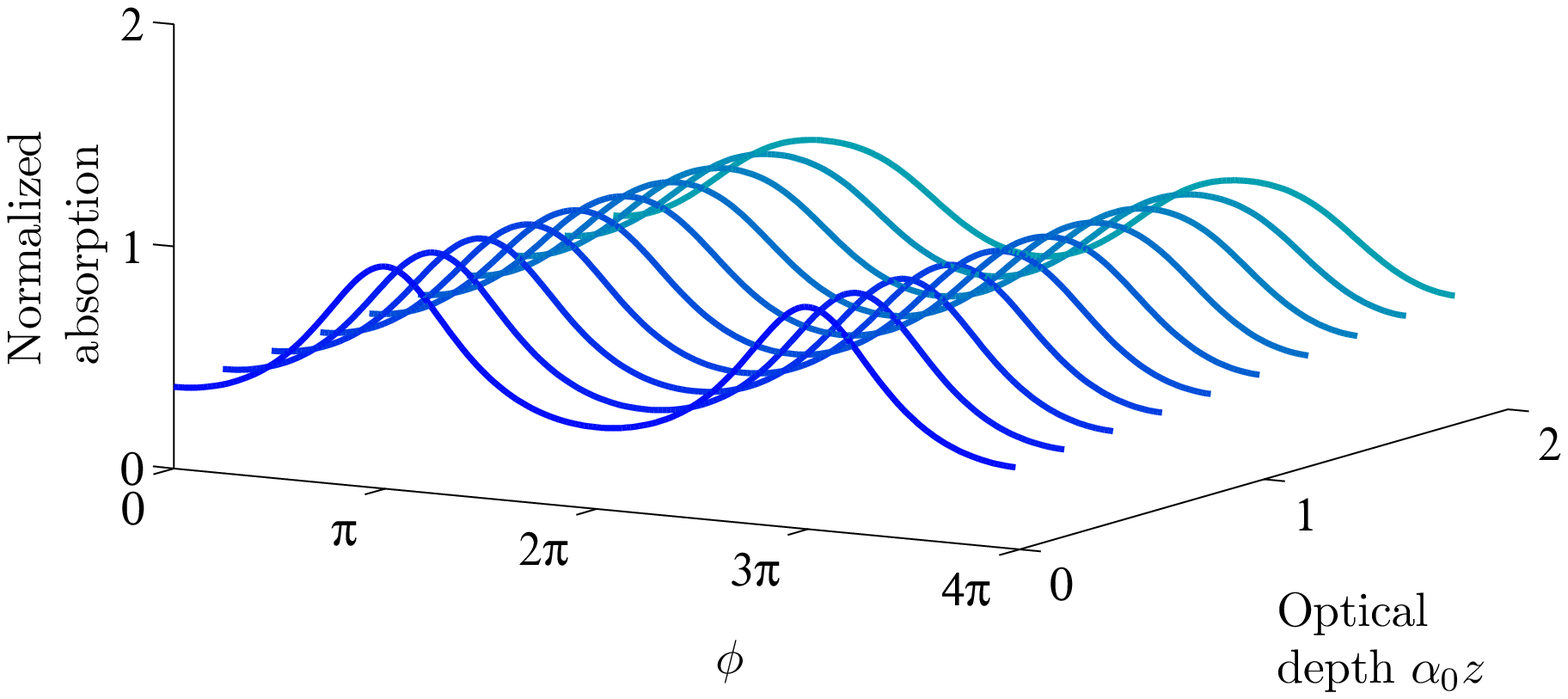}
\includegraphics[width=7.5cm]{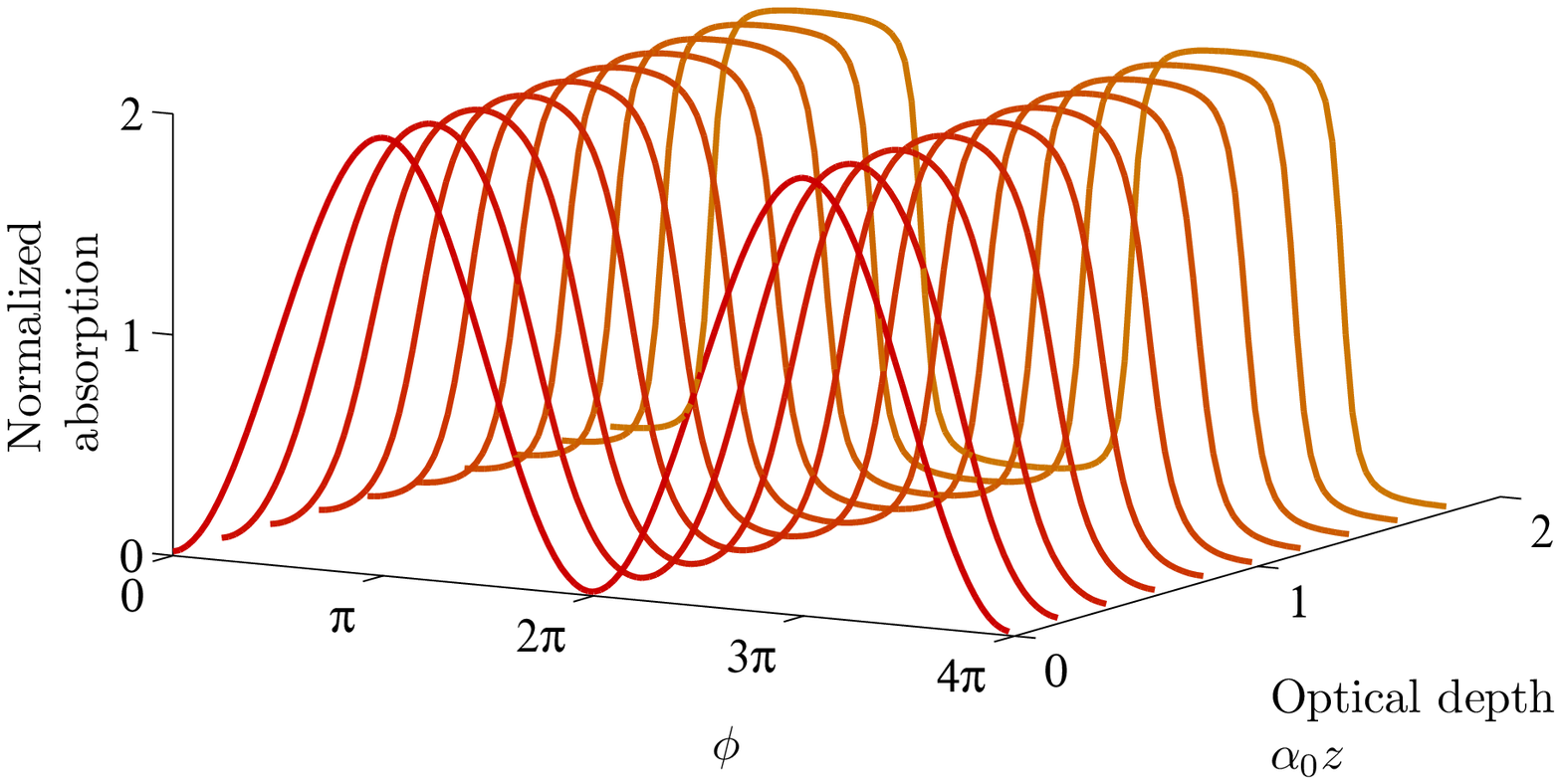}
\caption{(Color online) Numerical simulation of the evolution of a spectro-spatial grating in the absorption profile as the engraving fields propagate through the crystal in the case of a small angle between the beams. The optical pumping follows the standard scheme (top) or the \acro\ scheme (bottom). In the example shown here, $\zeta \avg{r}=0.9$ and $\xi \avg{r}=30$, respectively, corresponding to the maximum engraving power allowed in Tm:YAG crystals.}
\label{fig:profondeur_spectral}
\end{figure}

\subsection{Spectro-spatial grating with a small angle}
\label{sec:engraving_spectral}

We assume that the angle $\theta$ between the two engraving beams satisfies the condition $\theta\ll \sqrt{\frac{\lambda}{2L}}$ so that all diffraction orders may exist. The buildup of higher orders of diffraction along the propagation of the two engraving fields makes the excitation shape $r(\phi,z)$ evolve as the fields penetrate deeper into the atomic medium. As shown in Fig.~\ref{fig:profondeur_spectral}, the resulting absorption $\alpha(\phi,z)$ becomes progressively closer to a square profile.

This occurs for both the simple optical pumping scheme and the \acro\ scheme, with opposite consequences. In the former, because of the very restrictive weak field condition, the minimum absorption is always significant, and the engraving fields undergo absorption. Therefore the grating contrast (as defined in Eq.~\ref{eq:contrast}) decays as one gets deeper in the medium (from $0.63$ at the entrance to $0.41$ at the $\alpha_0L=2$ output if $\zeta \avg{r}=0.9$). In the latter, the propagation of the engraving beams through the spectro-spatial grating makes the grating evolve from a pure sinusoidal function with a contrast close to $2$, towards a square grating with the same average absorption and identical contrast. A square grating is known to give rise to the largest efficiency at a given average optical depth~\cite{bonarota2010efficiency}. One can therefore expect a low diffraction efficiency in the standard optical pumping scheme, and a high efficiency in the \acro\ scheme, larger than with a sinusoidal grating, but lower than with a square grating.

\begin{figure}[t]
\includegraphics[width=7.5cm]{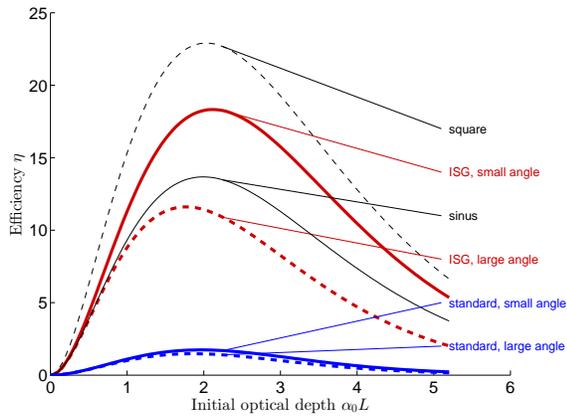}
\caption{(Color online) Maximum efficiencies in Tm:YAG calculated for different optical pumping schemes and beam configurations. The efficiencies calculated for uniform sinusoidal or square grating shapes as defined in Eq.~\ref{eq:sin} and \ref{eq:square} are also given.}
\label{fig:comparaison}
\end{figure}

We numerically simulate the propagation of a weak pulse propagating along $\textbf{k}_0$ in the non-uniform spectro-spatial grating described above. The first-order diffracted pulse is emitted along wavevector $\textbf{k}_1$ with a delay $\tau$. The diffraction efficiency $\eta$ is defined as the power ratio between the diffracted beam and the incoming beam (see appendix for details). The results are presented in Fig.~\ref{fig:comparaison}.
In the case of standard optical pumping in Tm:YAG, as expected, the diffraction efficiency is weak (below $1.75~\%$), because of the low grating contrast imposed by the weak field condition.
In the case of \acro, the weak field condition is much less restrictive and the grating shape is close to optimum: the efficiency reaches $\eta=18.3\%$ for an initial optical depth $\alpha_0 L=2$.

We also show in Fig.~\ref{fig:comparaison} the diffraction efficiencies calculated with two ideal spectro-spatial gratings with a contrast equal to $2$: a sinusoidal grating and a square grating, uniform over the whole atomic medium depth.
\begin{eqnarray}
\alpha_{sin}(\phi,z)&=&\alpha_0 (1+\sin\phi) \label{eq:sin}\\
\alpha_{sq}(\phi,z)&=&\alpha_0 (1+sign(\sin\phi)) \label{eq:square}
\end{eqnarray}
The sinusoidal grating leads to a $13.5\%$ maximum efficiency, whereas the square grating leads to $21.9\%$, for $\alpha_0L=2$. We verify that the efficiency with the \acro\ scheme in the small angle configuration lies between the sinusoidal and the square grating efficiencies.

\subsection{Spectro-spatial grating with a large angle}

\begin{figure}[t]
\includegraphics[width=7.5cm]{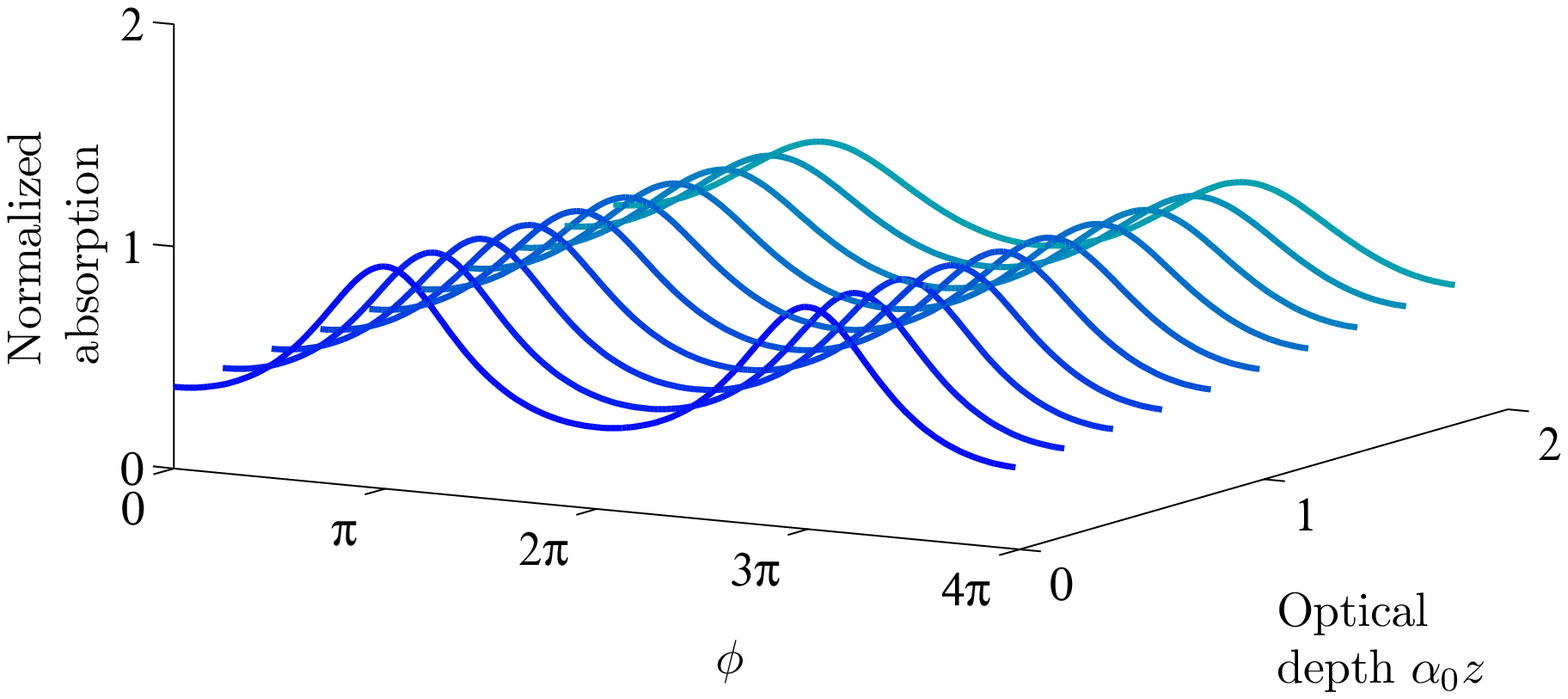}
\includegraphics[width=7.5cm]{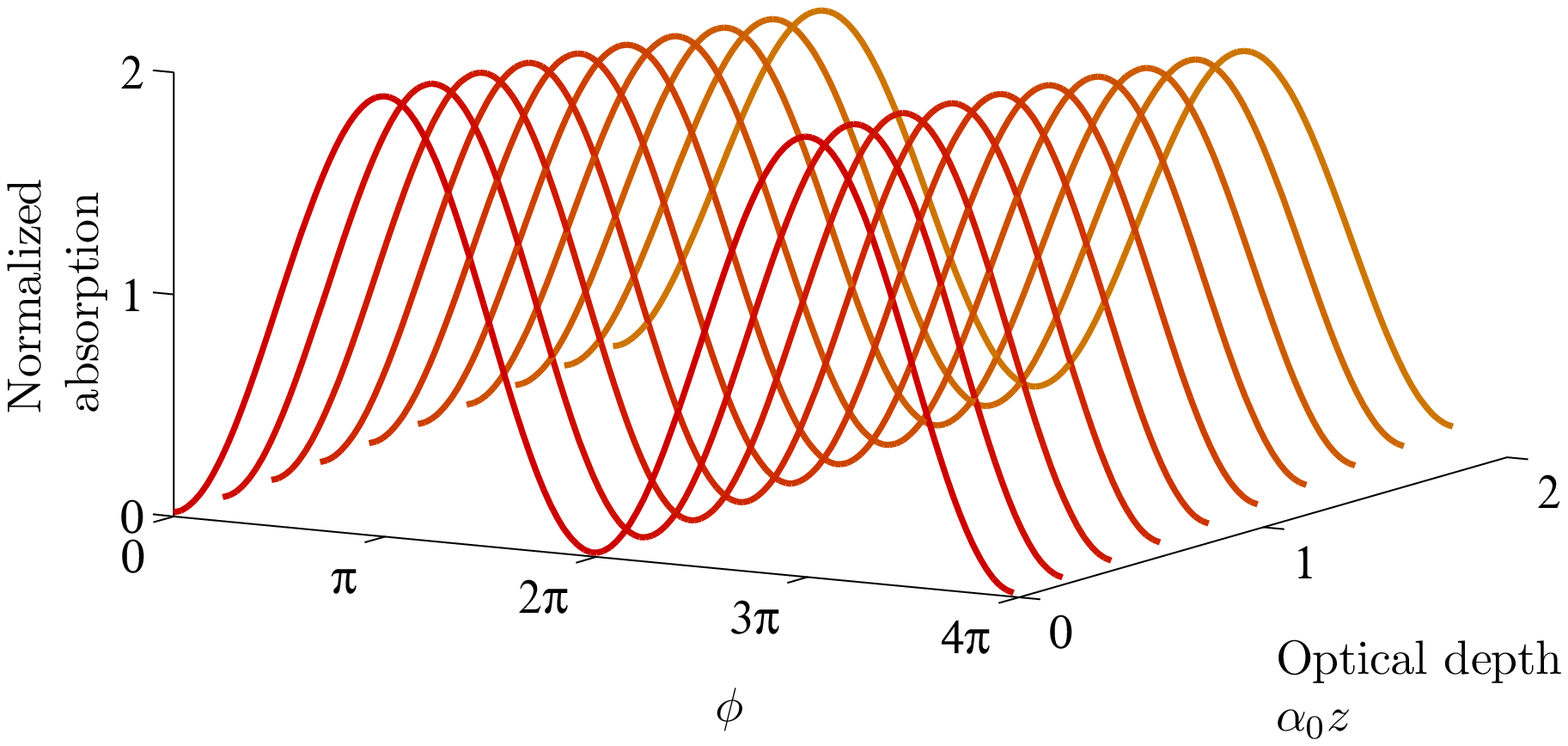}
\caption{(Color online) Numerical simulation of the evolution of a spectro-spatial grating in the absorption profile as the engraving fields propagate through the crystal in the case of a large angle between the beams. The optical pumping follows the standard scheme (top) or the \acro\ scheme (bottom). In the example shown here, $\zeta \avg{r}=0.9$ and $\xi \avg{r}=30$, respectively. }
\label{fig:profondeur_spectro-spatial}
\end{figure}

When the angle $\theta$ is larger than $\sqrt{\frac{\lambda}{2L}}$, the phase matching condition (Eq.~\ref{eq:PMC}) is violated for orders $n\geq2$. Only the two lowest orders (\emph{ie} the two incoming engraving pulses) can propagate and the excitation spectrum remains sinusoidal. As shown in Fig.~\ref{fig:profondeur_spectro-spatial}, in the two schemes, the grating contrast decays as the fields penetrate in the crystal, but its shape is not altered.
With standard optical pumping, the low initial grating contrast leads to the absorption of the engraving beams and a decay of the contrast (as defined in Eq.~\ref{eq:contrast}), from $0.63$ to $0.35$ after $\alpha_0L=2$. In the \acro\ scheme however, the decay is a little slower (from $1.97$ to $1.57$) since the minimum absorption is closer to zero. We still expect a good efficiency, although lower than the ideal sinusoidal grating efficiency.

The corresponding calculated diffraction efficiencies are plotted in Fig.~\ref{fig:comparaison}. Their behaviour is similar to the small angle configuration, only with lower values due to the less optimal grating shape. In the standard optical pumping scheme, the maximum available efficiency is only $1.5\%$. In the \acro\ configuration, the maximum efficiency occurs for $\alpha_0L=1.8$ and reaches $11.6\%$.

\section{Experimental demonstration}
\label{sec:exp}
\subsection{Experimental setup}

In our $2.5$-mm long, $0.5\%$ at. Tm-doped YAG crystal, we measure an initial optical depth $\alpha_0L=2$ at the center of the inhomogeneous line. A weak ($18$~G) magnetic field is applied along the $[001]$ crystalline axis with copper Helmholtz coils, to get $\Dge=500$~kHz in relevant, polarization-selected substitution sites.
The magnetic field splits the electronic levels into two nuclear sublevels due to thulium's $I=1/2$ nuclear spin. We consider that the level system dynamics behaves as depicted in Fig.~\ref{fig:5level}. The crystal is immersed in liquid helium at $2.2$~K. The homogeneous width along the optical transition is typically $10~$kHz in these conditions.

\begin{figure}[t]
\includegraphics[width=6cm]{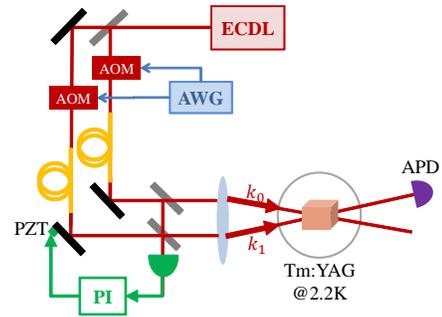}
\caption{(Color online) Schematic setup. ECDL: frequency-stabilized extended cavity diode laser; AWG: arbitrary waveform generator; AOM: acousto-optic modulator; PZT: piezo-electric transducer, PI: proportional-integral controller.}
\label{fig:setup}
\end{figure}

\begin{figure*}[t]
\includegraphics[width=13cm]{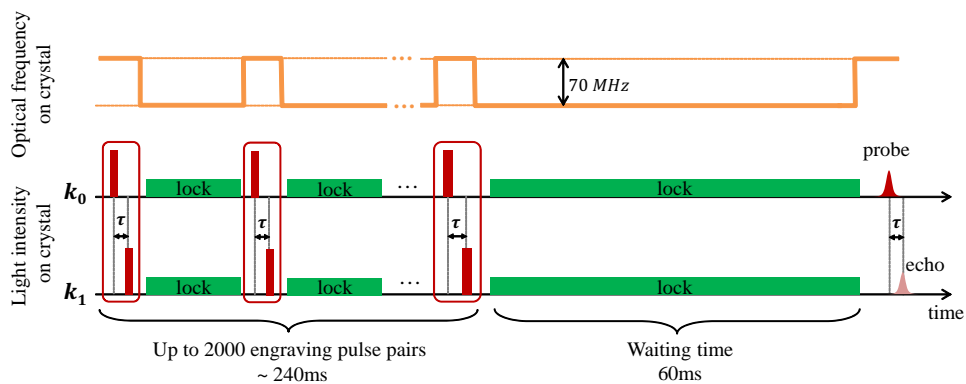}
\caption{(Color online) Experimental pulse sequence along wavevectors $\textbf{k}_0$ and $\textbf{k}_1$ used to generate an accumulated spectro-spatial grating. Frequency-detuned lock pulses are needed to ensure interferometric stability of the setup over several seconds.}
\label{fig:pulse_seq}
\end{figure*}

The optical setup is shown in Fig.~\ref{fig:setup} and the pulse sequence is shown in Fig.~\ref{fig:pulse_seq}. The grating is prepared by applying a sequence of rectangular pulse pairs, with duration $200$~ns, delay $\tau=1~\mu$s, area $0.013\pi$ and respective wavevectors $\textbf{k}_0$ and $\textbf{k}_1$, forming an angle $\theta$. The delay corresponds to a $1$~MHz grating period. The pairs are repeated every $120~\mu$s. Although the waiting time between two consecutive pulse pairs is small, the average pumping rate is close to $0.16 \gamma_e$. This way, the excitation obeys the weak field conditions given by equations~\ref{eq:weakR}.

The vibration isolation achieved by the optical table supports is not sufficient to maintain a constant relative phase between the two engraving pulses over the grating lifetime, \emph{ie} $5$~s. In order to dynamically adjust the path length difference, we generate quasi-continuous "lock pulses" between consecutive pulse pairs on both optical paths. These lock pulses are detuned from the engraving and probe pulses by $70$~MHz, recombined in the same spatial mode and sent to a PDA36A Thorlabs photodetector. The intensity on this detector is a sinusoidal function of the path length difference. The correction signal is generated with a Newport LB1005 P-I servo-controller and is fed to a piezoelectric actuator inserted in a standard mirror mount.
The lock pulses detuning is chosen such that the lock pulses lie outside our grating bandwidth, while still being accessible with standard, double-pass acousto-optical modulators.

After up to $2000$ pulse pairs, the engraving is stopped for $60$~ms to let atoms decay to the ground state sublevels.

A gaussian-shaped probe pulse is then sent along wavevector $\textbf{k}_0$.
The probe pulse is significantly longer (FWHM of $350$~ns) than the engraving pulses in order to probe the center of the engraved spectrum where the envelope of the grating can be regarded as constant. The echo is emitted along $\textbf{k}_0+\textbf{K}=\textbf{k}_1$, with delay $\tau$ from the probe pulse.
After the echo, the entire sequence starts over, with a $300$~ms cycle time.

All laser beams are generated with an extended cavity diode laser stabilized on a high finesse Fabry-Perot cavity with the Pound-Drever-Hall technique, ensuring the grating spectral stability. Two acousto-optic modulators in double pass setup modulate the intensity and frequency in the two beams, according to Fig.~\ref{fig:pulse_seq}, with the help of a dual-channel arbitrary waveform generator Tektronix AWG520.

%Magnetic field orientation gives a branching ratio $R=2\%$.

\subsection{Efficiency measured for small and large angles}

\begin{figure}[t]
\includegraphics[width=7.5cm]{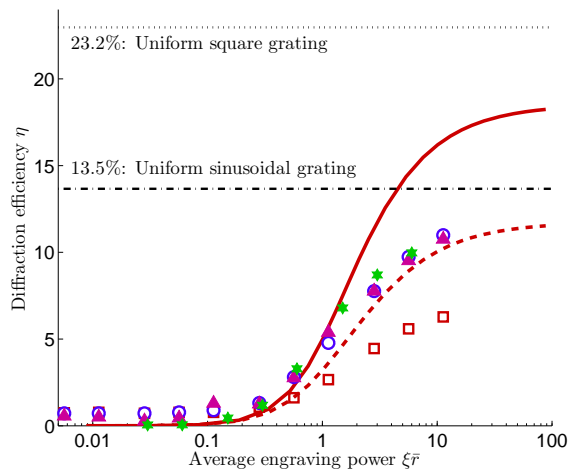}
\caption{(Color online) Diffraction efficiency with a spectral or spectro-spatial grating with storage in the Zeeman levels. Circles: pure spectral grating ($\theta=0$). Triangles and squares: spectro-spatial grating with $\theta=7.5$ and $17.5$~mrad, respectively.  Stars: efficiency calculated from the measured spectral gratings shown in Fig.~\ref{fig:zeeman5_allprofilexp}.
Thick solid line (resp., thick dashed line): numerical simulation for a small (resp., large) angle. Thin dashed-dotted line (resp., thin dotted line): calculated efficiency for a uniformly sinusoidal (respectively square) grating with an average optical depth of $2$.}
\label{fig:zeeman5_exp}
\end{figure}

The $L=2.5$~mm crystal length, and optical wavelength $\lambda=793$~nm lead to the small angle condition for the two engraving pulses: $\theta\ll12.6$~mrad.
We measure the diffraction efficiency for three different angles: $\theta=0$, $\theta=7.5$~mrad and $\theta=17.5$~mrad, which correspond respectively to a spectral grating, a spectro-spatial grating with small angle, and a spectro-spatial grating with large angle. The experimental data are presented in Fig.~\ref{fig:zeeman5_exp}, together with the calculated diffraction efficiency with $\alpha_0 L=2$. The average engraving power $\xi\avg{r}$ is varied between $0.005$ and $11.3$ by spanning the number of pulse pairs per cycle from $1$ to $2000$.

We observe identical results in the zero and small angle configuration, which agrees with the discussion in Sec.~\ref{sec:deep}. As in the simulation, the experimental efficiency grows with engraving power. The maximum efficiency in that case is $11~\%$, instead of $16.5\%$ as expected from the numerical simulation. When the angle is larger, the efficiency is reduced, and the maximum efficiency drops to $6.3~\%$ instead of $10.3\%$. For much larger angles, the efficiency should remain the same, provided the engraving beams perfectly overlap.

The experimental data qualitatively agree with the simulations. The quantitative discrepancy could come from an inefficient engraving process, or from an inefficient diffraction process. To determine which one is responsible for this discrepancy, we directly measure the shape of the absorption grating, in the case of a spectral grating.

\begin{figure}[t]
\includegraphics[width=7.5cm]{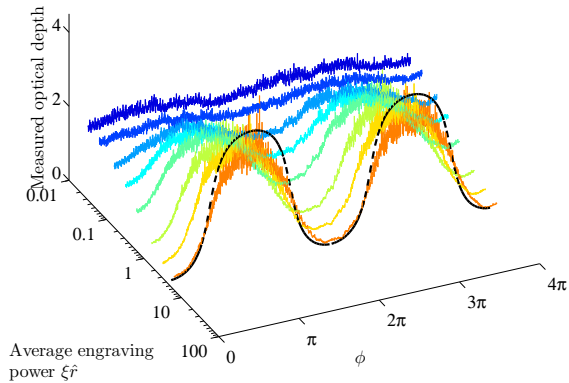}
\caption{(Color online) Absorption profiles averaged over the sample depth. Colored solid lines: measured with a weak chirped probe. Black dashed line: calculated, for $\xi\avg{r}=6$.}
\label{fig:zeeman5_allprofilexp}
\end{figure}

\subsection{Direct measurement of the absorption profile}
After having built a purely spectral grating with collinear engraving beams, we measure the transmission of a weak probe whose frequency is linearly swept over $3$~MHz in $200~\mu$s. With this measurement we can derive the grating profile for different average engraving powers (see Fig.~\ref{fig:zeeman5_allprofilexp}).

The absorption profile is expected to be non-uniform throughout the sample.
With this experiment, we are only able to measure the absorption averaged over the crystal depth. We compare with the calculated average absorption profile for $\xi\avg{r}=6$ corresponding to the curve with largest engraving power. Although the calculated and experimental profiles have a similar shape, the experimental contrast ($c_{exp}=1.60$) is lower than expected ($c_{calc}=1.82$).
In our simulation, we disregarded the homogeneous linewidth that should give rise to a smoother absorption profile. What's more, our model assumes that the weak transitions ($\ket{1} \rightarrow \ket{4}$ and $\ket{3} \rightarrow \ket{2}$) are forbidden, although in our setup, their oscillator strength is only $7$ times weaker than that of the strong transitions. These two simplifications probably explain the contrast discrepancy between the simulation and the experiment.
%other reasons: residual mechanical instabilities

We calculate the diffraction efficiency that one would expect from such a grating, assuming it is uniform throughout the sample depth. We plot these results on Fig.~\ref{fig:zeeman5_exp}, together with the efficiencies measured by comparing the intensities of the probe and first diffraction order fields. The two methods give exactly the same efficiencies, proving that our model describes well the diffraction process.

In this demonstration, the grating bandwidth is rather low ($3$~MHz) but already exceeds the Zeeman splitting in the ground state: $\Delta_g=600$~kHz. It can be increased with no fundamental limitation.

\section{Conclusion}
Interlaced Spin Grating is a scheme for the preparation of spectro-spatial gratings in a inhomogeneously broadened absorption profile. It relies on the storage of atoms in two long-lived ground state sublevels, leading to low power requirements. Its bandwidth is not limited by the ground state splitting.
In depth investigation of the optical pumping dynamics specific to \acro\ and beam propagation has enabled us to identify the best conditions to get large efficiency, via an enhancement of the grating contrast, together with quasi-optimal grating shape and preservation of the optical depth. We have shown that the efficiency could reach $18.3\%$ in the small angle configuration and $11.6\%$ with a large angle, which is about $10$ times larger than observed in the first experimental demonstrations of broadband ISG~\cite{bonarota2011highly,saglamyurek2011}.
We were able to demonstrate $11$\% experimental diffraction efficiency with a zero or small angle, and $6.3$\% with a large angle.

The \acro\ scheme is only suitable when dealing with periodic gratings, and not for gratings with variable spacing (such as the dispersive line used for time reversal~\cite{linget2013} or the pulse sequencing architecture~\cite{saglamyurek2014}).
It can prove particularly useful in spectro-spatial holography architectures such as the rainbow analyzer~\cite{lavielle2003}.
%In this application, a spectro-spatial grating is created over the whole inhomogeneous bandwidth of Tm:YAG ($20$~GHz), with angularly and frequency-swept engraving beams. Using the long-lived nuclear spin sublevels in the \acro\ configuration would allow for a long accumulation time, hence a reduction of the necessary power, and a large diffraction efficiency, leading to a better dynamic range for the rainbow analyzer.

\section*{Acknowledgements}
This work is supported by the french national grant RAMACO no. ANR-12-BS08-0015-02. The research leading to these results has received funding from the People Programme (Marie Curie Actions) of the European Union's Seventh Framework Programme FP7/2007-2013/ under REA grant agreement no. 287252.

\appendix
\section{Propagation of two fields in an absorbing medium}
\label{appendix}

\subsection{Propagation of the engraving fields}
\label{app:propag}
Let us consider two light pulses with a time delay $\tau$ with wavevectors $\textbf{k}_0$ and $\textbf{k}_1$, illuminating an inhomogeneously broadened medium. At the front of the sample ($z=0$), the two combined engraving fields are described by the following spectral amplitude:
\begin{equation}
E(\textbf{x},z = 0, \phi) = |E_0| e^{-i\textbf{k}_0\cdot \textbf{x}}(1+ e^{-i \phi})
\end{equation}
where $\textbf{x}$ is the transverse atomic position vector in the plane perpendicular to the propagation direction, and $\phi=2\pi\nu \tau+\textbf{K}\cdot\textbf{x}$ is the spectro-spatial phase, and $\textbf{K}=\textbf{k}_1-\textbf{k}_0$.
The pumping rate $R(z,\phi)$ is proportional to the field power spectral intensity $|E(z,\phi)|^2$.

The propagation of the engraving field $E$ is described by the wave equation
\begin{equation}
\Delta E +k^2 \epsilon_r E=0
\label{eq:waveeq}
\end{equation}
where $k=\left|\textbf{k}_0\right|=\left|\textbf{k}_1\right|$, the electric relative permittivity $\epsilon_r$ is related to the absorption coefficient $\alpha$ via the Kramers-Kronig relation: $\epsilon_r=1-\frac{i}{k} (1+i \mathcal{H})\alpha$, and $\mathcal{H}$ is the Hilbert transform. Assuming the absorption coefficient is a $2 \pi$-periodic function of $\phi$, we write it as its Fourier expansion:
\begin{equation}
\alpha^{(p)}(z)=\frac{1}{2\pi} \int_{<2\pi>} \alpha (z,\phi)e^{i p \phi} d\phi
\end{equation}
Similarly, the field $E$ can be written into multiple spectro-spatial modes as:
\begin{equation}
E^{(p)}(z)=\frac{1}{2\pi} \int_{<2\pi>} E(z,\phi)e^{i p \phi} d\phi
\end{equation}
$E^{(0)}$ and $E^{(1)}$ are the two incoming engraving fields.
From Eq.~\ref{eq:waveeq} we derive the coupled wave equations for mode $p$:
\begin{eqnarray}
\frac{\partial E^{(p)}(z)}{\partial z}+\frac{\alpha^{(0)}(z)}{2} E^{(p)}(z)-i \frac{p(p-1)K^2}{2 k} E^{(p)}(z)\\=-\sum_{q>0} \alpha^{(q)}(z) E^{(p-q)}(z)
\label{eq:coupled_angle}
\end{eqnarray}
These equations show that higher orders $E^{(p)}$ with $p\geq2$ are created from lower orders as the fields propagate through the medium.

%Condition for a small angle between the beams.
The third term in Eq.~\ref{eq:coupled_angle} accounts for the phase shift that builds up between the spectro-spatial mode $E^{(p)}$ and the periodic structure that generates it. For $p=0$ and $p=1$, corresponding to the incoming engraving fields, this phase shift is zero. Higher-order modes only exist under the condition of a small angle $\theta$ between $k_0$ and $k_1$, that satisfies $\theta\ll\sqrt{\frac{\lambda}{2L}}$, where $L$ is the medium length along the $z$ direction.

\begin{itemize}
\item In the case of a small angle between the two engraving beams $\theta\ll\sqrt{\frac{\lambda}{2L}}$, all orders of diffraction exist. Eq.~\ref{eq:coupled_angle} reduces to :
    \begin{equation}
\frac{\partial \left|E(z,\phi)\right|^2}{\partial z}+\alpha(z,\phi)\left|E(z,\phi)\right|^2 =0
\label{eq:intensity_small}
\end{equation}

\item In the case of a large angle between the two engraving beams, only orders $0$ and $1$ may propagate:
    \begin{eqnarray}
    \frac{\partial E^{(0)}(z)}{dz}+\frac{\alpha^{(0)}(z)}{2}E^{(0)}(z) &=& 0\\
    \frac{d E^{(1)}(z)}{d z}+\frac{\alpha^{(0)}(z)}{2}E^{(1)}(z) &=& -\alpha^{(1)}(z)E^{(0)}(z)
    \end{eqnarray}
\end{itemize}
%When solving these equations, one must keep in mind that $\alpha$ is a function of $z$ and $\phi$, given by Eq.~\ref{eq:alpha3} or \ref{eq:alpha3Z}.

\subsection{Propagation of probing field; efficiency}
\label{app:probing}
To calculate the diffraction efficiency of the grating, we consider the propagation of the probe field $\EE^{(0)}$.
The engraved absorption profile is a function of depth $z$, transverse position in the crystal $\textbf{x}$ and frequency $\nu$. The diffraction efficiency after a distance $L$ in the crystal is given by:
\begin{equation}
\eta=\left| \frac{\EE^{(1)}(L)}{\EE^{(0)}(0)} \right|^2
\end{equation}
The diffracted field $\EE^{(1)}$ is solution of
\begin{eqnarray}
\frac{d \EE^{(0)}(z)}{d z}+\frac{\alpha^{(0)}(z)}{2}\EE^{(0)}(z) &=& 0\\
\frac{d \EE^{(1)}(z)}{d z}+\frac{\alpha^{(0)}(z)}{2}\EE^{(1)}(z)&=& -\alpha^{(1)}(z)\EE^{(0)}\!(z)
\end{eqnarray}
with the initial condition: $\EE^{(1)}(0)=0$.

Note that for a uniformly engraved absorption profile, $\alpha^{(0)}$ and $\alpha^{(1)}$ are independent of the medium depth. The diffraction efficiency is given by:
\begin{equation}
\eta=\left( \alpha^{(1)} L \right)^2 e^{-\alpha^{(0)}L}
\end{equation}

%%%%%%%%%%%%%%%%%%%%%%% References %%%%%%%%%%%%%%%%%%%%%%%%%
%\bibliography{SpectralGrating}

\end{document}